\documentclass[prl,amsmath,twocolumn,superscriptaddress]{revtex4-1}

\usepackage[utf8]{inputenc}
\usepackage{graphicx}
\usepackage{dcolumn}
\usepackage{bm}
\usepackage{textcomp}  
\usepackage{siunitx}
\usepackage{braket}
\usepackage{commath}
\usepackage{parskip}
\usepackage{graphicx}
\usepackage{xcolor}
\usepackage{amsfonts}
\usepackage{tabularx}
\usepackage{array}

\usepackage{wrapfig}

\begin{document}
\title{Intermediate Field Coupling of Single Epitaxial Quantum Dots to Plasmonic Waveguides}

\author{Michael Seidel}
\affiliation{Experimental Physics III, University of Bayreuth, 95447 Bayreuth, Germany}

\author{Yuhui Yang} 
\affiliation{Institute of Solid State Physics, Technische Universität Berlin, 10623 Berlin, Germany}

\author{Thorsten Schumacher}
\affiliation{Experimental Physics III, University of Bayreuth, 95447 Bayreuth, Germany}

\author{Yongheng Huo} 
\affiliation{Institute of Semiconductor and Solid State Physics, Johannes Kepler University Linz, Altenbergerstra{\ss}e 69, A-4040 Linz, Austria}

\author{Saimon Filipe Covre da Silva}
\affiliation{Institute of Semiconductor and Solid State Physics, Johannes Kepler University Linz, Altenbergerstra{\ss}e 69, A-4040 Linz, Austria}

\author{Sven Rodt}
\affiliation{Institute of Solid State Physics, Technische Universität Berlin, 10623 Berlin, Germany}

\author{Armando Rastelli}
\affiliation{Institute of Semiconductor and Solid State Physics, Johannes Kepler University Linz, Altenbergerstra{\ss}e 69, A-4040 Linz, Austria}

\author{Stephan Reitzenstein}
\email{stephan.reitzenstein@physik.tu-berlin.de}
\affiliation{Institute of Solid State Physics, Technische Universität Berlin, 10623 Berlin, Germany}

\author{Markus Lippitz}
\email{markus.lippitz@uni-bayreuth.de}
\affiliation{Experimental Physics III, University of Bayreuth, 95447 Bayreuth, Germany}

\date{\today}



\begin{abstract}
\begin{wrapfigure}[16]{r}{82.5mm}
\hspace*{-30mm}\includegraphics[width=82.5mm]{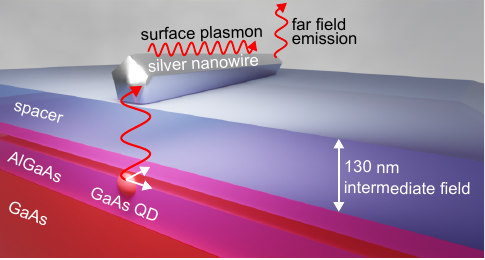}
\end{wrapfigure}

Key requirements for quantum plasmonic nanocircuits are reliable single-photon sources, high coupling efficiency to the plasmonic structures and low propagation losses. Self-assembled epitaxially grown GaAs quantum dots are close to ideal stable, bright and narrowband single-photon emitters. Likewise, wet-chemically grown monocrystalline silver nanowires are among the best plasmonic waveguides. However, large propagation losses of surface plasmons on the high-index GaAs substrate prevent their direct combination. Here, we show by experiment and simulation that the best overall performance of the quantum plasmonic nanocircuit based on these building blocks is achieved in the intermediate field regime with an additional spacer layer between the quantum dot and the plasmonic waveguide. High-resolution cathodoluminescence measurements allow a precise determination of the coupling distance and support a simple analytical model to explain the overall performance. 
The coupling efficiency is increased up to four times by standing wave interference near the end of the waveguide.

\end{abstract}

\maketitle

Quantum photonics has the potential to revolutionize our world with breakthrough technologies such as quantum computing and quantum communication, for instance using quantum dots (QDs) as single photon emitters~\cite{Heindel23}. Especially in terms of applications, scalability is indispensable, and integrated photonic networks are highly sought after~\cite{Moody2022}. Plasmonic nanocircuits are a promising platform since they not only dramatically reduce circuit size, but also allow light to be controlled and manipulated at a truly nanoscale level~\cite{Bozhevolnyi2017a,Xu2018a,Fernandez-Dominguez2018,Bogdanov2019,Schoerner2020}.
Even though many electrons are involved in the surface plasmon polariton (SPP), the quantum-optical nature is preserved~\cite{Dlamini2018}.

In recent years, the coupling of various quantum emitters to plasmonic waveguides has been demonstrated (for a review see~\cite{Kumar2021}). 
Sources of single plasmons have been reported at both room and liquid helium temperatures using various combinations of waveguides and \mbox{emitters~\cite{Akimov2007,Kolesov2009,Huck2011,Kumar2013,Li2015,Kress2015,Wu2017,Li2018,Kumar2018,Kumar2019,Grandi2019}}. However, for true quantum-optical operation of the circuit, high-quality sources of single photons are essential. Epitaxially grown self-assembled quantum dots possess close to ideal quantum properties as they are bright, non-blinking, and have narrow \mbox{linewidths~\cite{Somaschi2016,zhou2023epitaxial,Tomm2021}}.
Nonetheless, the typical approach of bringing the waveguide close to the emitter does not work, as the high refractive index of the semiconductor host induces significant damping of SPP due to ohmic and radiative losses. Additionally, the optical properties of epitaxial quantum dots degrade with decreasing distance from the dot's surface~\cite{heyn2017droplet}. These issues have previously been addressed through an indirect coupling method using dielectric-plasmonic mode conversion~\cite{Wu2017}, requiring significant nanofabrication.

\begin{figure}[t!]
\centering
\includegraphics[scale=1]{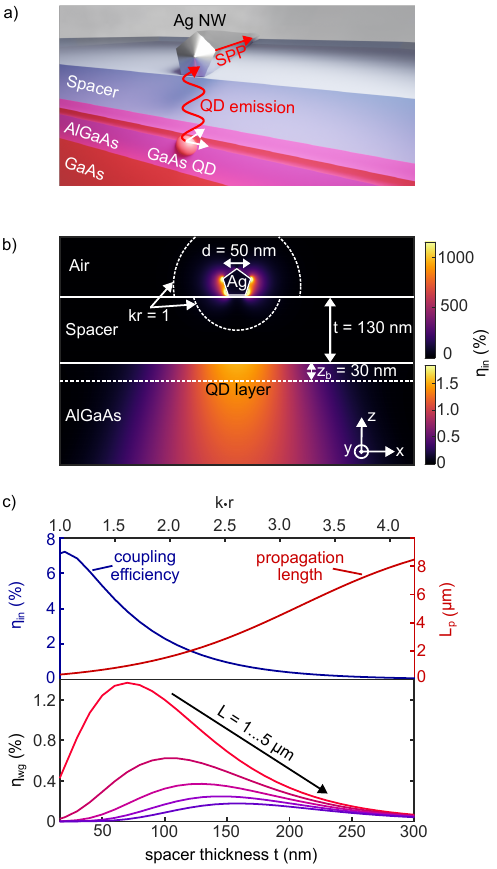}
\caption{\textbf{Intermediate field coupling of a quantum dot to a plasmonic waveguide.}
a)~Sketch of the coupling scheme: A quantum dot embedded within AlGaAs barriers radiatively couples to a silver nanowire that is separated by the capping layer and an additional dielectric spacer.
b)~Spatial variation of the coupling efficiency $\eta_{in}$ into the waveguide for an emitter located in the \mbox{xz-plane} and with dipole moment in the \mbox{xy-plane} of the sample. The circle enclosing \mbox{$kr=1$} illustrates the transition between near and far field. Note the different color scale bars for the lower and upper half spaces.
c)~Upper panel: Coupling efficiency $\eta_{in}$ and propagation length $L_{p}$, as a function of the spacer thickness $t$. The coupling efficiency is evaluated for an emitter that is centrally located beneath the nanowire in the quantum dot layer and placed at a depth $z_{b}=30$\,nm below the spacer/AlGaAs interface.
Lower panel: waveguide efficiency $\eta_{wg}$ as a function of spacer \mbox{thickness $t$} and waveguide length $L$. Optimal performance is achieved in the intermediate field for $kr \gtrsim 1$.}
\label{fig:decay_rate_wg}
\end{figure}

Here, we address conflicting requirements in a different and much simpler way: Instead of placing the plasmonic waveguide directly on the semiconductor substrate containing quantum dots, a planar dielectric layer with a lower refractive index is used as a spacer between the semiconductor host and plasmonic waveguide, as depicted in Fig.~\ref{fig:decay_rate_wg}a. This allows us to balance the efficiency of coupling $\eta_{in}$ and propagation $\eta_{p}$: a thicker spacer layer enhances the plasmon propagation length, a thinner layer enhances the coupling efficiency between the quantum dot and the waveguide. In the following, we demonstrate that the nanocircuit performance is expected to be superior when the coupling between the emitter and waveguide occurs in the intermediate field ($kr\approx1$), rather than the near field ($kr\ll1$), where $k$ and $r$ denote the light wavevector and radial distance from the emitter, respectively.

As sketched in Fig.~\ref{fig:decay_rate_wg}a, we assume an infinitely extended waveguide in propagation direction. Mode profiles and corresponding effective mode indices are calculated as a function of the spacer thickness $t$ with Comsol Multiphysics. For all simulations we use $\lambda = 790$\,nm,  the emission wavelength of our GaAs quantum dots~\cite{DaSilva2021}. The chemically grown monocrystalline silver nanowire~\cite{song2011} is modeled with a pentagonal cross-section ($d=50$\,nm).  The refractive indices for silver ($n_{Ag}=0.035+5.49i$) and AlGaAs ($n_{AlGaAs}=3.44$) are taken from literature \cite{Johnson1972,Aspnes1986}. The dielectric spacer (spin-on-glass IC1-200 from Futurrex) is modeled with $n_{spacer}=1.41$ according to the manufacturer. 

To compute the coupling efficiency $\eta_{in}$ for the plasmonic waveguide mode, we follow the framework of Ref.~\cite{Chen2010}. The decay rate of the emitter into the plasmonic mode is related to the dot product between its transition dipole moment $\boldsymbol{\mu}$ and the modal field $\boldsymbol{E}^{mode}$. We have to consider that the transition dipole moments of our quantum dots are given by two energetically almost degenerate exciton states oriented orthogonal to each other in the sample plane. For the sake of simplicity, we assume that one of the dipole moments is oriented parallel and the other one perpendicular to the nanowire axis. The coupling efficiency
\begin{equation}
\eta_{in} = \sum_{j} \left |\boldsymbol{\mu}_{j} \cdot \boldsymbol{E}^{mode}(x,z) \right |^2
\label{eqn:eta_in_xz}
\end{equation}
is then obtained by incoherently adding up the two dipole moment contributions $j=x,y$ and normalizing to the emission of a dipole in homogeneous AlGaAs (see Supporting Information S1).

The coupling efficiency $\eta_{in}$ for a spacer thickness of \mbox{$t=130$\,nm} is shown in Fig.~\ref{fig:decay_rate_wg}b as a function of the emitter position in the \mbox{$xz$-plane}.
Note that in the simulation we place the dipole not only within the AlGaAs matrix as in the experiment, but also inside the dielectric spacer and around the nanowire. Apart from the strongly confined hot spots directly at the nanowire, there is a less confined region in the semiconductor where the incoupling efficiency $\eta_{in}$ does not vary much. At the quantum dot burial depth $z_{b}\approx30$\,nm, $\eta_{in}$ drops only by a factor of two when leaving the wire axis laterally by 125\,nm. This stems from the rather loosely bound character of the waveguide mode and relaxes the required QD alignment accuracy, even though the absolute coupling efficiency is lower. Furthermore, the weak depth dependence suggests that the QD can be placed deeper in the AlGaAs without much change in coupling efficiency. This is particularly interesting when considering that the optical properties of the QD improve rapidly with increasing burial depths~\cite{heyn2017droplet}.

The effect of the spacer thickness on the waveguide coupling efficiency $\eta_{in}$ and the propagation length $L_{p}$ is shown in the top panel of Fig~\ref{fig:decay_rate_wg}c: With increasing layer thickness, the amplitude of the waveguide mode at the quantum dot position is reduced, and therefore the coupling efficiency decreases. Here we evaluate the coupling efficiency for an emitter that is centered with respect to the nanowire, at a depth of $z_{b}=30$\,nm. On the other hand, the propagation length $L_{p}$ of the mode is strongly increased for thicker spacers. This is due to the diminishing influence of the high-index AlGaAs, resulting in a mode that more strongly bound to the waveguide and features less radiative losses.

For the experimental realization of our plasmonic coupling concept, we are interested in the waveguide efficiency $\eta_{wg}=\eta_{in} \, \eta_{p}$, which also includes the propagation efficiency $\eta_{p}=e^{-L/L_{p}}$ for a waveguide of finite length $L$. Obviously, the optimal spacer thickness also depends on the waveguide length $L$, as can be seen in the lower panel of Fig.~\ref{fig:decay_rate_wg}c. Accordingly, the highest waveguide efficiency $\eta_{wg}$ is achieved for short waveguides and rather thin spacers. For an experimentally meaningful nanocircuit, however, the waveguide should be longer than the spatial resolution of the optical microscope, i.e. $L \gtrsim 1$\,µm. For such waveguide lengths we find the optimum in the intermediate field regime at $kr=1.6-2.6$ or $t=70-160$\,nm. Here, the spacer thickness $t$ is rewritten in terms of $kr=k_{0}(n_{AlGaAs}z_{b}+n_{spacer}t)$ with the vacuum wavevector $k_{0}$. The transition from near to far field is also illustrated as a circle enclosing $kr=1$ in Fig.~\ref{fig:decay_rate_wg}b.

Let us now turn to the experimental realization of such an intermediate field coupling. The sample is based on near-surface self-assembled GaAs quantum dots in \mbox{AlGaAs} barriers grown by molecular beam epitaxy on a GaAs substrate. For the dielectric spacer, the polysiloxane-based spin-on glass (\mbox{IC1-200} Intermediate Coating, Futurrex) is spin-coated on top of the GaAs surface, resulting in a film with a thickness of \mbox{$t = (130\pm15)$\,nm}. Chemically grown monocrystalline silver nanowires (PL-AgW100, PlasmaChem) with average widths of \mbox{$d=(50\pm10)$\,nm} and typical lengths of a few micrometers are dispersed on top of the IC1 film. A detailed description of the sample fabrication can be found in the Supporting Information S2. The random arrangement of dots and wires samples all relative orientations and coupling distances, requiring preselection of potentially coupled quantum dot -- nanowire pairs. Therefore, we determine the spatial arrangement by high-resolution cathodoluminescence mapping and then measure the waveguide performance by optical microscopy.

\begin{figure*}
\centering
\includegraphics[scale=1]{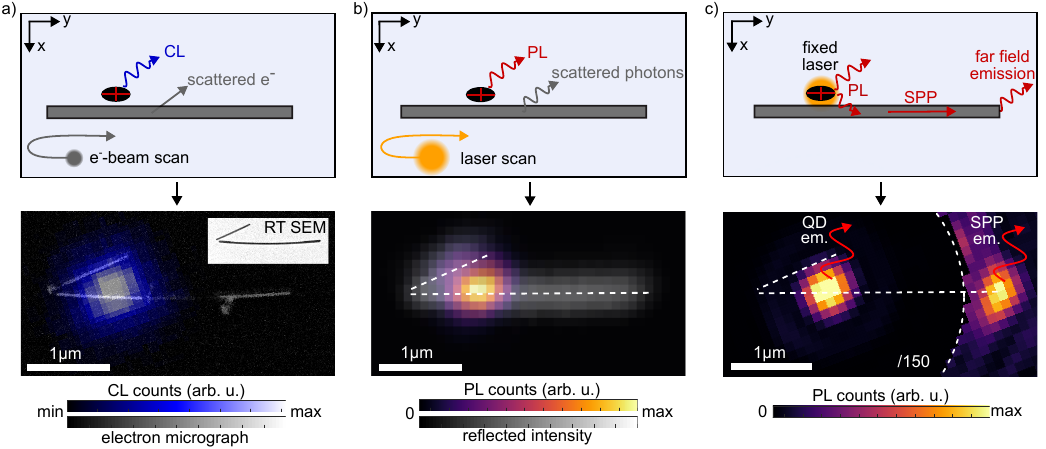}
\caption{\textbf{Investigation of a coupled quantum dot - nanowire system by complementary methods.} 
a)~Raster-scanning the electron beam while detecting the cathodoluminescence and scattered electrons. The obtained cathodoluminescence image with overlaid electron micrograph allows a precise distance measurement. The inset depicts a room temperature scanning electron micrograph of the same nanowire.
b)~Sketch of confocal laser scanning imaging with corresponding photoluminescence data, overlaid with the reflection image of the structure.
c)~Proof of intermediate field coupling by plasmon propagation imaging via a CCD-camera and stationary excitation of the quantum dot. In the recorded image, the area around the quantum dot is software-attenuated by a factor of 150 to increase the visibility.}
\label{fig:SPG}
\end{figure*}

Low-temperature cathodoluminescence combines high-resolution electron microscopy with access to quantum dot emission, making it an excellent technique to specify the relative positions of quantum dots and nanowires. Our setup is described in detail in Ref.~\cite{Rodt21}. The sample is mounted on a liquid He-flow cryostat (20\,K) and excited with a focused electron beam of 20\,kV acceleration voltage, which is scanned over the sample surface. As sketched in Fig.~\ref{fig:SPG}a, the cathodoluminescence emission of each excitation spot position is collected by a spectrometer, simultaneously providing a secondary electron image and cathodoluminescence spectrum mapping with the same coordinates.

An example of a data set can be found in the lower section of Fig.~\ref{fig:SPG}a. For comparison, a room temperature scanning electron micrograph is included as an inset. Although the diameter of the electron beam is only a few nanometers, the actual size of the cathodoluminescence spots mostly results from the effective diameter of the generation volume and charge carrier diffusion \cite{Davidson1980,Yacobi1986}. Hence, a two-dimensional Gaussian profile is used to fit the cathodoluminescence emission spots, allowing for a precise determination of the relative lateral positions of quantum dots and nanowires with $10-30$\,nm accuracy.
For the depicted nanosystem, the cathodoluminescence image reveals the QD positions $x_{QD}=(77\pm12)$\,nm and $y_{QD}=(685\pm26)$\,nm with respect to the nanowire end.
However, electron beam scanning is not suitable to distinguish between direct QD emission and remote plasmon emission at the nanowire end due to lacking spatial resolution in the detection path. 

Consequently, we use an all-optical confocal microscope to demonstrate intermediate field coupling. A fast scan mirror moves the excitation laser focus (\mbox{NA $=0.9$}, 635\,nm wavelength) over the sample inside a closed-cycle cryostat (20\,K). Different detection schemes are employed. To identify the preselected QD-nanowire system, we map the sample by photon counting and a combination of photoluminescence and reflection (Fig.~\ref{fig:SPG}b). In order to enhance the contrast in the reflection image, the direct laser reflection is suppressed with a polarizer. Slight sample drifts during the laser scans can be neglected since this measurement is only used for identification of the nanowires.

Intermediate field coupling is demonstrated in Fig.~\ref{fig:SPG}c by launching and detecting plasmons: Here, the excitation laser is stationary focused on the QD while the surrounding sample area's luminescence is imaged onto a CCD-camera. We observe clear emission from the SPP that is launched by the coupled quantum dot and scattered at the nanowire's end. We find identical photoluminescence spectra for the direct QD emission and the outcoupled photons of the plasmon (see Supporting Information~S3). Emission from the short wire end is also expected but is experimentally hidden in the airy-patterned background of the direct QD emission (see Supporting Information~S4).

We analyzed a total of nine QD-nanowire systems, which differ by up to a factor of 80 in the intensity ratio of the respective SPP emission $I_{pl}$ and the direct QD emission $I_{qd}$ (see Supporting Information Tab.~S3). In the following, we extract the coupling efficiency for these nanosystems and explain this --~on the first sight~ -- large variation as an interference effect near a waveguide end.

\begin{figure}
\centering
\includegraphics[scale=1]{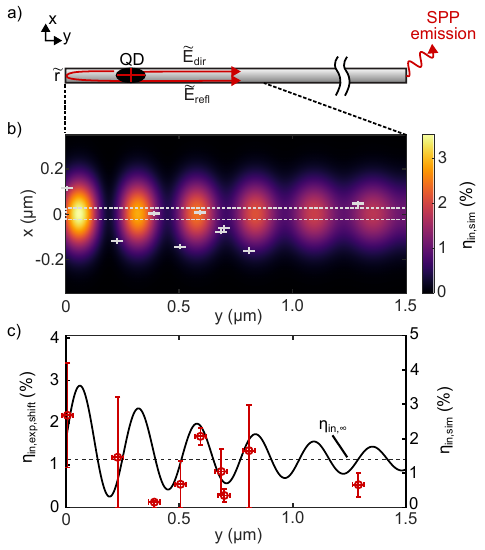}
\caption{\textbf{Interference caused by reflection at the waveguide end explains the spatial variation of the incoupling efficiency.}
a)~Schematic of the interfering SPPs, arising from non-zero reflectivity at the nanowire termination.
b)~Coupling efficiency map in the sample plane at the burial depth \mbox{$z_{b}=30$\,nm}. The lateral quantum dot positions are displayed by gray crosses, the size of which indicates the position uncertainty from the cathodoluminescence images. The dashed horizontal lines correspond to the nanowire width.
c)~Simulated coupling efficiency (black line) along the nanowire axis and experimental coupling efficiency (red circles) according to Eq.~\ref{eqn:eta_in_exp_shift} and corrected for the offset $x_{QD}$ from the nanowire axis. The dashed line represents the coupling efficiency $\eta_{in,\inf}\approx 1.4\,\%$ for a quantum dot centered below an infinitely extended wire (see Fig.~\ref{fig:decay_rate_wg}).}
\label{fig:model}
\end{figure}

All nine QD-nanowire systems are formed by QDs near (about~1\,\textmu m) one end of the silver waveguide. Considering typical propagation lengths in the range of a few micrometers and the small diameter of our silver nanowires, we expect substantial reflection of the SPP at the near wire end~\cite{Kolesov2009}. This results in an interference \mbox{$|E|^2=|E_\text{dir}+E_\text{refl}|^2$} of the direct surface plasmon $E_\text{dir}$ and the reflected surface plasmon $E_\text{refl}$ (see Fig.~\ref{fig:model}a) and consequently in a position-dependent coupling efficiency. In the worst case, both fields destructively interfere with each other, and no net coupling would be observable, although the emitter is close to the nanowire.

To model the position-dependent coupling efficiency in the \mbox{$xy$-plane}, we make use of reciprocity and assume a semi-infinite wire. The mode profile in the \mbox{$xz$-plane} is already shown in \mbox{Fig.~\ref{fig:decay_rate_wg}b}. We keep the $z$ coordinate constant at the burial depth \mbox{$z_{b}=30$\,nm} to obtain the mode profile $\boldsymbol{E}^{mode}(x)$. In propagation direction ($y$), we interfere the direct wave and the reflected wave, both propagating with an effective mode index $\tilde{n}_\text{eff}$. The reflection coefficient of the wire end is also complex-valued $\tilde{r}=r e^{i\phi_{r}}$ with the reflection amplitude $r$ and reflection phase $\phi_{r}$. Overall, we obtain the coupling efficiency in the $xy$-plane
\begin{equation}
\eta_{in,sim}(x,y) = \sum_{j} \left| E^{mode}_{j}(x) \left( 1 + \tilde{r} \,  e^{2 \, i \, k_{0} \, \tilde{n}_\text{eff} \, y} \right) 
\right|^2
\label{eqn:interference}
\end{equation}
by incoherently summing the dipole moment contributions in $j=x,y$.
The resulting coupling efficiency map (Fig.~\ref{fig:model}b) shows the expected oscillatory interference features that decay
in interference contrast
with distance to the near waveguide end, as the amplitudes of $E_\text{dir}$ and $E_\text{refl}$ separate.
Although measured at different waveguides, we draw all nine investigated structures in this map by overlaying the nanowire ends, which already suggests strong fluctuations in their coupling efficiency.

In the experiment, the photon rate detected at the outcoupling end of the waveguide is given by the product of the partial efficiencies for incoupling, propagation, outcoupling and detection times the QD's bare emission rate. At the QD position, we detect this bare rate times the QD detection efficiency. Knowing all these factors from either numerical simulations or measurements (see Supporting Information S5) allows us to calculate back to the experimentally observed incoupling efficiency $\eta_{in,exp}$ at the specific QD positions relative to the waveguide.

For comparison with the interference model, it is more convenient to compare a one-dimensional data set. Knowing the spatial mode profile, we shift the experimental QD positions to below the waveguide ($x=0$\,nm) by
\begin{equation}
\eta_{in,exp,shift}=\eta_{in,exp} \;  \frac{\sum_{j} |E^{mode}_{j}(0)|^2}{\sum_{j} |E^{mode}_{j}(x_{QD})|^2}
\label{eqn:eta_in_exp_shift}
\end{equation}
for the offset $x_{QD}$ of the respective QD via the mode profile $E^{mode}(x)$. This is the incoupling efficiency that would have been measured if the dot had been centered below the waveguide. Fig.~\ref{fig:model}c compares these values with the model at $x=0$\,nm. We have fixed the reflection amplitude ($r=0.65$) and phase ($\phi_{r}=-\pi/2$), as well as the air-sided far-field collection efficiency ratio ($\eta_{spp,ff} / \eta_{qd,ff} = 17$) based on our numerical simulations (see Supporting Information S5), and vary mode index, propagation length and an overall scaling parameter. The optimal fit is achieved with a mode index of $n_\text{eff}=1.53$, a propagation length $L_{p}=0.86$\,µm and a scaling factor of $1.23$. These values are already used to plot Fig.~\ref{fig:model}b.

We find good agreement between our interference model and the corrected coupling efficiency, although not all individual variations of the nanosystems are taken into account. Minor differences in geometry parameters can shift the datapoints somewhat. In particular, imperfections such as slightly bent wires or small kinks can cause additional losses due to reflections or far field scattering but were not observed in the propagation images and are therefore neglected. Furthermore, the exact dipole moment orientations within each QD are unknown to us, which only affects QDs located far away from the nanowire axis (see Supporting Information S6). The resulting uncertainty can be quantified and is included in the error bars for the experimental coupling efficiency. In addition, the errorbars comprise the uncertainty arising from the the QD-SPP emission ratio extraction, and the uncertainty in the lateral QD position determination.

Nevertheless, the fit parameters consistently lie within a realistic range. For the mode index, we expect values ranging from $n_{\text{eff}}=1.5-1.7$ from numerical simulations, depending on the details of the chosen geometry. The propagation length in the finite element simulation (see Fig.~\ref{fig:decay_rate_wg}c) is approximately $L_{p}=2-3$\,µm, around three times greater than the fit outcome, which is attributed to material imperfections and residual surfactant at the nanowire's surface. We experimentally extracted a propagation length of $L_{p}\approx1.0$\,µm from laser transmission experiments, which is consistent with the fit result but also subject to large variations (see Supporting Information~S7). Additionally, we find an overall scaling parameter of $1.23$, indicating that all major contributions are included in the model.

The interference model in Fig.~\ref{fig:model}c implies that the waveguide coupling rate is substantially larger toward the near end of the wire compared to the infinite length waveguide. The coupling rate can in principle be increased up to a factor of four for a reflection coefficient of $r=1$. This would result in a coupling efficiency of $5.5\,\%$. The overall efficiency of the device could be further optimized by impedance-matching the waveguide ends~\cite{Grimm2021} and achieving constructive interference with substrate reflections~\cite{Wu2017}.

In summary, we have demonstrated the coupling of single self-assembled GaAs quantum dots to silver nanowires in the intermediate field. This is achieved by balancing coupling and propagation efficiency, using a planar dielectric spacer of about 130\,nm thickness. The relative positions of quantum dots and nanowires are determined with high accuracy by simultaneously imaging them through low-temperature cathodoluminescence. This enabled us to establish an interference model that explains the varying coupling efficiencies. The reflection of the propagating plasmon at the wire's near end boosts the efficiency by up to four times. Intermediate field coupling does not necessitate nanostructuring processes in the QD's dielectric surroundings, which often degrade the (quantum) optical characteristics of the QD. Furthermore, the intermediate field approach is not limited to QDs grown near the surface because of its weak depth dependence (see Fig.~\ref{fig:decay_rate_wg}b). Taken together, this means that a Fourier-limited source of single plasmons is within reach.\\

\begin{acknowledgments}
Acknowledgements: This work was funded by the German Research Foundation (INST 131/795-1 320 FUGG, INST 91/310-1 FUGG), European Union’s Horizon 2020 Research and innovation Programme under the Marie Sklodowska-Curie Grant Agreement No. 861097 (QUDOT-TECH), Einstein foundation via the Einstein Research Unit “Perspectives of a quantum digital transformation: Near-term quantum computational devices and quantum processors” and the Austrian Science Fund (FWF) via the Research Group FG5, I 4320, I 4380

\end{acknowledgments}

\bibliography{references}

\clearpage

\onecolumngrid
\appendix

\section{Supporting Information: Intermediate Field Coupling of Single Epitaxial Quantum Dots to Plasmonic Waveguides}

\author{Michael Seidel}
\affiliation{Experimental Physics III, University of Bayreuth, 95447 Bayreuth, Germany}

\author{Yuhui Yang} 
\affiliation{Institute of Solid State Physics, Technische Universität Berlin, 10623 Berlin, Germany}

\author{Thorsten Schumacher}
\affiliation{Experimental Physics III, University of Bayreuth, 95447 Bayreuth, Germany}

\author{Yongheng Huo} 
\affiliation{Institute of Semiconductor and Solid State Physics, Johannes Kepler University Linz, Altenbergerstra{\ss}e 69, A-4040 Linz, Austria}

\author{Saimon Filipe Covre da Silva}
\affiliation{Institute of Semiconductor and Solid State Physics, Johannes Kepler University Linz, Altenbergerstra{\ss}e 69, A-4040 Linz, Austria}

\author{Sven Rodt}
\affiliation{Institute of Solid State Physics, Technische Universität Berlin, 10623 Berlin, Germany}

\author{Armando Rastelli}
\affiliation{Institute of Semiconductor and Solid State Physics, Johannes Kepler University Linz, Altenbergerstra{\ss}e 69, A-4040 Linz, Austria}

\author{Stephan Reitzenstein}
\email{stephan.reitzenstein@physik.tu-berlin.de}
\affiliation{Institute of Solid State Physics, Technische Universität Berlin, 10623 Berlin, Germany}

\author{Markus Lippitz}
\email{markus.lippitz@uni-bayreuth.de}
\affiliation{Experimental Physics III, University of Bayreuth, 95447 Bayreuth, Germany}

\date{\today}

\maketitle

\section{S1 - Definition of the coupling efficiency}

Based on the framework in Ref.~\cite{Chen2010}, we define the coupling efficiency
\begin{equation}
\eta_{in}(x,z)=
\frac{3 \pi c \epsilon_{0} \boldsymbol{E}(x,z) \cdot \boldsymbol{E}^{*}(x,z)}
{n_{AlGaAs}\,k_{0}^{2} \int_{A} S_{y}\,dA}
\end{equation}
with the free-space speed of light $c$, the vacuum permittivity $\epsilon_{0}$, the modal electric field~$\boldsymbol{E}$ in the transverse $xz$-plane, the refractive index $n_{AlGaAs}=3.44$, the vacuum wavevector~$k_{0}$, and the time-averaged Poynting vector component $S_{y}=\frac{1}{2}\Re{(\boldsymbol{E} \times \boldsymbol{H}^{*})} \cdot \boldsymbol{y}$ in waveguide direction~$y$, which is integrated over the transverse plane $A$. The coupling efficiency $\eta_{in}$ is normalized to the emission of a dipole in homogeneous AlGaAs, and therefore can exceed unity.\\

Furthermore, it is practical to define the normalization constant
\begin{equation}
p=\frac{3 \pi c \epsilon_{0}}{n_{AlGaAs} k_{0}^{2} \int_{A} S_{y}\,dA},
\end{equation}
to obtain the normalized electric field
\begin{equation}
\boldsymbol{E}_{norm}(x,y,z)=\frac{\boldsymbol{E}(x,y,z)}{\sqrt{p}}.
\end{equation}
Now the coupling efficiency directly follows from
\begin{equation}
\eta_{in}(x,y,z)=\boldsymbol{E}_{norm}(x,y,z) \cdot \boldsymbol{E}_{norm}^{*}(x,y,z).
\end{equation}
The normalized fields $\boldsymbol{E}_{norm}$ allow to evolve the coupling efficiency along the waveguide axis by combining the 2D mode analysis with Eq.~2 in the main text, as done in Fig.~3b in the main text. Considering the case $r=0$, we find a constant coupling efficiency along the y-axis, as one would expect without reflections.


\section{S2 - Sample fabrication}

The GaAs/AlGaAs quantum dot (QD) samples used in this work are grown after the recipe described in Ref.~\cite{Huo2013} and have also been utilized in Ref.~\cite{Zhang2015, Wu2017}. Specifically, three slightly different samples with QD burial depths $z_{b}=$\,15\,nm, 30\,nm and 40\,nm were used. The influence of the burial depth on the waveguide coupling efficiency is negligible according to Fig.~1b in the main text.

For the dielectric spacer we use a polysiloxane-based spin-on glass (IC1-200 Intermediate Coating, Futurrex). IC1-200 is diluted in isobutanol with a 1:1 mixture. Afterwards, the IC1 is spin-coated on top of the semiconductor at 77\,rps and baked out on a hotplate at 200\,°C for two minutes, resulting in a film thickness of around 130\,nm, which is confirmed by AFM measurements.

Finally, chemically-grown monocrystalline silver nanowires (PL-AgW100, diluted in isopropanol, PlasmaChem) are dispersed on top of the IC1 film. Afterwards, the sample is rinsed gently in ethanol in order to remove the PVP (Polyvinylpyrrolidone) surfactant, and dried in nitrogen flux. In order to avoid degradation of the Ag nanowires (NWs), the samples are coated with 5\,nm of Al\textsubscript{2}O\textsubscript{3} by atomic layer deposition. During transport and storing, the samples are kept under vacuum conditions.

\clearpage

\section{S3 - Additional data for coupled QD-NW systems and methods}

We have collected complete datasets for nine coupled quantum dot - nanowire (QD-NW) systems. A dataset (measured at 20\,K if not noted else) comprises:\\

- \textbf{Cathodoluminescence / Secondary electron microscope (SEM) images}\\
- \textbf{Photoluminescence / Reflection images via confocal laser scanning}\\
- \textbf{Waveguide propagation images}\\
- \textbf{QD and SPP emission spectra}\\

\begin{figure}[h]
\centering
\includegraphics[scale=1]{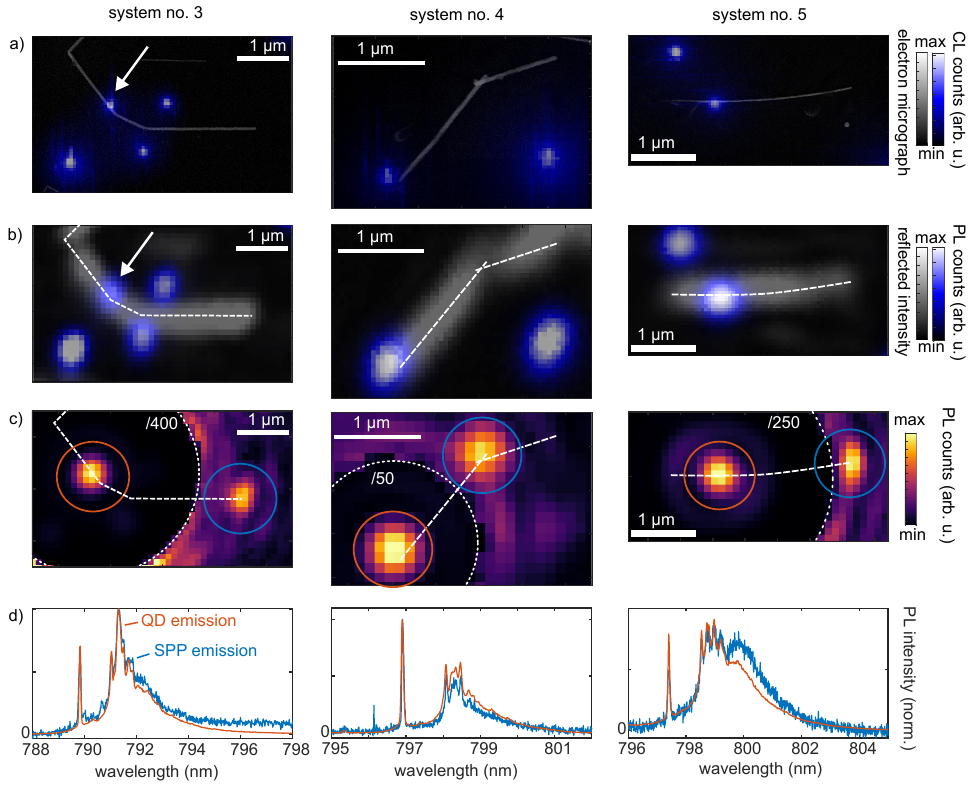}
\caption{\textbf{Additional data on three coupled QD-NW systems.}
a)~Overlaid cathodoluminescence and SEM scans.
b)~Overlaid photoluminescence and polarized reflection scans.
c)~Propagation image, demonstrating launching of SPPs. The red circle indicates the excited QD, the blue circle highlights the scattered SPP at the nanowire end. The region around the excited QD is software attenuated to increase visibility.
d)~Photoluminescence spectra of the QD emission and the outcoupled SPP.}
\label{fig:all_wires_SI}
\end{figure}

In Fig.~\ref{fig:all_wires_SI}, complete datasets for three QD-NW systems are shown, including QD emission and SPP spectra which are not shown in the main text. Tab.~S3 gives an overview of the experimental quantitites extracted from these measurements. In the following, we discuss applied methods and data processing in greater detail.\\

\begin{table*}[h]
{\def\arraystretch{1}\tabcolsep=10pt
\begin{tabularx}{0.9\textwidth}{p{5cm}*{9}{>{\centering\arraybackslash}X}}
& \multicolumn{9}{c}{system no.} \\
& 1 & 2 & 3 & 4 & 5 & 6 & 7 & 8 & 9 \\
\hline
NW diameter (nm)            & 48 & 53 & 53 & 62 & 42 & 60 & 63 & 50 & 45 \\
NW length (µm)              & 2.87 & 3.75 & 4.25 & 1.63 & 2.87 & 2.61 & 5.05 & 3.71 & 3.27 \\
QD burial depth (nm)        & 15 & 30 & 40 & 40 & 40 & 30 & 30 & 30 & 30 \\
IC1 thickness (nm)			& 131 & 162 & 136 & 136 & 136 & 162 & 162 & 162 & 162 \\
$x_{QD}$ (nm)               & 77 & 8 & 48 & 115 & 62 & 162 & 118 & 4 & 143 \\
$y_{QD}$ (nm)               & 685 & 593 & 1290 & 6 & 698 & 807 & 228 & 390 & 505 \\
$I_{spp}/I_{qd}~(\cdot 10^{-3}$) & 5.4 & 4.2 & 1.6 & 17.7 & 2.1 & 5.0 & 0.22 & 0.26 & 0.85 \\
$\eta_{in,exp}~(\%)$ & 0.66 & 1.68 & 0.49 & 1.21 & 0.24 & 0.41 & 0.63 & 0.13 & 0.22 \\
$\eta_{in,exp,shift}~(\%)$ & 0.85 & 1.68 & 0.54 & 2.18 & 0.29 & 1.34 & 1.19 & 0.13 & 0.55 \\
\end{tabularx}
}
\caption{\textbf{Overview of the extracted experimental quantities for nine coupled QD-NW-systems.} Nanowire dimensions and relative QD positions ($x_{QD}$, $y_{QD}$) are obtained from cathodoluminescence/SEM images. The emission ratio $I_{spp}/I_{qd}$ is obtained from the \mbox{waveguide} propagation images. For the extraction of the coupling efficiency $\eta_{in,exp}$ and $\eta_{in,exp,shift}$, see Supporting Information S5. QD burial depth and IC1 film thickness are given for the sake of completeness.}
\label{tab:tab}
\end{table*}


\textbf{Cathodoluminescence / SEM images}\\
From the cathodoluminescence/SEM images (Fig.~2a in the main text and Fig.~\ref{fig:all_wires_SI}a) we extract the quantum dot positions $x_{QD}$ and $y_{QD}$ relative to the nanowire end, with the y-axis parallel and the x-axis perpendicular to the nanowire axis. This is done by fitting a background-corrected two-dimensional Gaussian function to the cathodoluminescence spot of the respective quantum dot. The uncertainties in $x_{QD}$ ($y_{QD}$) shown in Fig.~3b,c in the main text are composed of the uncertainty of the fit and the determination of the nanowire center (end). As the SEM image and the corresponding cathodoluminescence map are recorded simultaneously and therefore share the same coordinates, these uncertainties are small ($<30$\,nm).\\

\textbf{Photoluminescence / Reflection via confocal laser scanning}\\
As mentioned in the main text, we perform two subsequent confocal laser scans and detect photoluminescence and reflection. Afterwards, both images are overlaid by transparency (Fig.~2b in the main text and Fig.~\ref{fig:all_wires_SI}b). For the reflection mapping, polarisation contrast is utilized to enhance the visibility of the silver nanowires. Therefore, the reflected laser (\mbox{$\lambda=635\,$nm}) is suppressed with an analyzer. As a result, only light which is polarized along the nanowires is collected by the APD based single-photon counting module. For a single scan, the contrast depends on the direction of the nanowires. Consequently, we add up several scans for different laser polarizations. This allows us to map the optical (photoluminescence/reflection) images to the cathodoluminescence/SEM images. Thereby, we rotate and scale the optical axes according to coordinate system given by the cathodoluminescence, which we expect to be the most accurate.


\textbf{Waveguide propagation imaging}\\
For the demonstration of QD-NW coupling, the QD is optically excited and the surrounding including the Ag nanowire is imaged onto the CCD-camera, while the excitation laser (\mbox{$\lambda=635\,$nm}) is filtered out by a bandpass. The scattered SPP signal is detected together with the direct QD emission (Fig.~2c in the main text and Fig.~\ref{fig:all_wires_SI}c). A background-corrected two-dimensional elliptical Gaussian is fitted to the quantum dot emission and the out-coupled surface plasmon emission, respectively. Integration of the Gaussian function finally leads to the SPP-QD-emission ratio $I_{spp}/I_{qd}$. This emission ratio will be used to determine the coupling efficiency $\eta_{in,exp}$ for each nanosystem in Supporting Information~S5.\\
For a subset of the investigated QD-NW systems, no clear SPP emission is observed. Instead of omitting these datasets, we decided to include these into our model (Fig.~3, main text). Just like the other systems, a 2D Gaussian is fitted at the nanowire end where the outcoupling is expected. The resulting SPP-QD emission ratio is declared as an upper limit for the true signal. Consequently, for these datapoints the uncertainty bar goes down to zero in Fig.~3c in the main text. As can be seen in the same figure, the $x_{QD}$-corrected coupling efficiency of these datapoints is consistent with our model. In other words, the vanishing emission is explained by a small coupling efficiency at the respective QD position, either due to large lateral offset from the wire axis, or destructive interference of the SPPs.\\

\textbf{QD and SPP emission spectra}\\
The spectra of the direct QD photoluminescence and the scattered SPP (Fig.~\ref{fig:all_wires_SI}d) are recorded by spatial filtering before entering the entrance slit of the spectrometer. We find identical spectra for direct QD emission and remote SPP emission.


\section{S4 - Signal-to-Background ratio} 

\begin{figure}[h]
\centering
\includegraphics[scale=1]{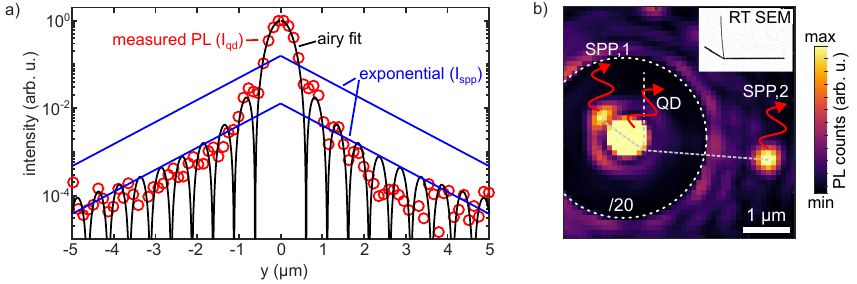}
\caption{\textbf{Dependence of the SPP-emission to QD-background on the length of the nanowire.}
a) Measured (red circles) and fitted (black line) intensity cross section of an uncoupled QD emission, imaged onto the CCD-camera. For comparison, the blue lines represent the expected SPP emission intensities for the nanosystems with the highest and lowest coupling efficiency as function of the distance from the emitter at $y=0$.
b) QD-NW coupling image of nanosystem~no.~2. Clear SPP emission is observable at the far wire end (SPP, 2), while the SPP emission at the near wire end (SPP, 1) is overlapping with the QD's airy pattern (QD). Another short wire does not show SPP emission. Inset: room-temperature SEM image of the same nanosystem.}
\label{fig:airy_fit}
\end{figure}

For most of the nanosystems, emission is observed only from the far wire end. For the near wire end, which typically is in a distance below 1\,µm from the QD, the SPP emission competes with a much stronger airy-patterned background of the direct QD emission, as shown in Fig.~\ref{fig:airy_fit}a. Here, the intensity cross section of an uncoupled QD emission image is fitted with an airy function.
We compare the QD intensity distribution with an exponential function which represents the expected SPP emission intensity (following Eq.~\ref{eqn:eta_in_exp}, see Supporting Information S5)
\begin{equation}
I_{spp}(y) = I_{qd}(0)\,\eta_{in} (1-|r|^2) \frac{\eta_{spp,ff}}{\eta_{qd,ff}} e^{-y/L_{p}}
\end{equation}
as a function of the distance $y$ from the QD. Here, the QD intensity $I_{qd}(0)=1$ is normalized and the other parameters $r=0.65$, $L_{p}=0.86$\,µm and $\frac{\eta_{spp,ff}}{\eta_{qd,ff}}=17$ are taken as in the main text. For the coupling efficiency $\eta_{in}$, we use the values for the nanosystems with highest and lowest efficiency according to Tab.~S3, respectively. It can be seen that for small distances, the airy-patterned background dominates, matching our observations. In this simple picture, different coupling efficiencies shift the exponential function against the background.

These findings are supported by nanosystem no. 2, featuring the highest coupling efficiency (see Tab.~S3) and signal-to-background ratio. Fig.~\ref{fig:airy_fit}b shows the QD-NW coupling image for this nanosystem with SPP emission from the short wire end (SPP,~1), overlapping with the QD's airy pattern (QD). At the far wire end (SPP,~2), the SPP-signal to QD-background ratio is much higher. Another short wire does not show any SPP emission, which can be explained by the larger distance from the QD and the large angle with the longer wire.

\clearpage

\section{S5 - Extraction of the coupling efficiency}
We write the detected intensity at the QD ($I_{qd}$) or waveguide end ($I_{spp}$) as product of the respective efficiencies, starting from the same total QD emission $I_{0}$. For the direct QD signal, only the far-field collection efficiency  $\eta_{qd,ff}$ enters. For the waveguide, we need to take into account the coupling efficiency $\eta_{in}$ and the plasmon propagation length. Moreover, the finite reflection $r$ at the waveguide end and the far-field collection efficiency  $\eta_{spp,ff}$ need to be considered. All together we get
\begin{equation}
\frac{I_{spp}}{I_{qd}} = \frac{I_{0} \eta_{in} \eta_{p} (1-|r|^2) \eta_{spp,ff}}{I_{0} \eta_{qd,ff}}
\end{equation}
with the field reflection amplitude $r$ of the wire end, effectively reducing the number of out-coupled photons. Furthermore, we account for different far-field collection efficiencies for the surface plasmon $\eta_{spp,ff}$ and the quantum dot $\eta_{qd,ff}$. This leads to the experimental coupling efficiency
\begin{equation}
\eta_{in,exp} = \frac{I_{spp}/I_{qd}}{e^{-L/L_{p}} (1-|r|^2) \frac{\eta_{spp,ff}}{\eta_{qd,ff}}}.
\label{eqn:eta_in_exp}
\end{equation}
The QD-SPP emission ratio is extracted from the waveguide propagation images, as discussed in Supporting Information S3. The propagation losses are corrected for the length~$L$, defined as the distance from the QD to the far wire end, which is known from the cathodoluminescence/SEM images. The propagation length $L_{p}$ is a free parameter in the interference model (Fig.~3, main text), where it is determined to $L_{p}=0.86$\,µm. This leaves the wire end reflection amplitude $r$ and the far-field collection efficiencies $\eta_{spp,ff}$ and $\eta_{qd,ff}$, which will be derived from simulations in the following.\\

\begin{figure}[b]
\centering
\includegraphics[scale=1]{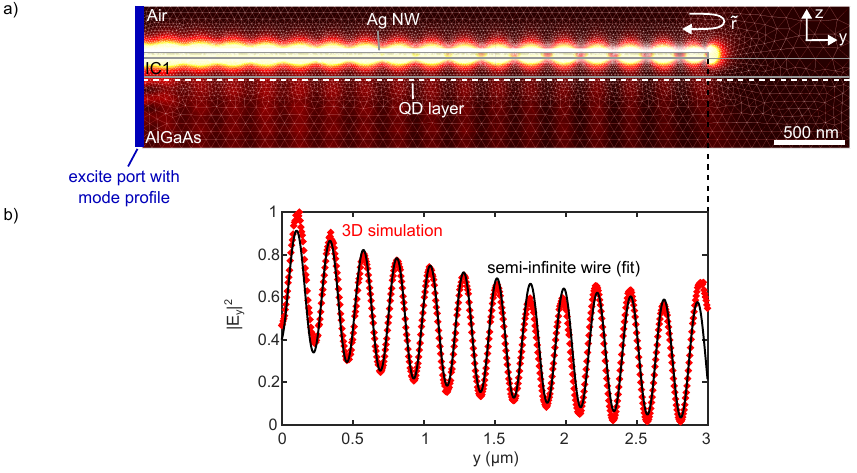}
\caption{\textbf{Extraction of the complex reflection coefficient for our geometry.}
a) A 3D model including a terminated nanowire is excited with the mode profile from Fig.~1b in the main text. The structure is shown from the side.
b) Comparison of analytic model (black line) and numerical model (red dots). The 3D port simulation is evaluated at a centered linecut at the QD layer in a depth $z_b=30$\,nm.}
\label{fig:port}
\end{figure}

\textbf{Complex reflection coefficient}\\
In order to determine the complex reflection coefficient $\tilde{r}=re^{i\phi_{r}}$ for our system, a 3D finite element simulation (Comsol Multiphysics) is performed. The cross section of the 3D model is identical to the geometry shown in Fig.~1b in the main text, but the nanowire is terminated before reaching the end of the computation window, as sketched in Fig.~\ref{fig:port}a. We take the mode profile in the $xz$-plane and excite the 3D model via the port function in Comsol to obtain the electric field distribution, which is evaluated in a centered linecut at $z_b=30$\,nm (with respect to the nanowire axis).
Now, we fit the analytical intensity distribution for the semi-infinite wire
\begin{equation}
|E(y)|^2 = \left | E_{0} \left [ e^{i \tilde{\beta} y} + e^{i \tilde{\beta}  (y_{0}-y)} \tilde{r} e^{i \tilde{\beta}  y_{0}} \right] \right |^2
\end{equation}
to the simulated intensity, featuring the expected interference fringes (Fig.~\ref{fig:port}b). Here, $E_{0}$ is the initial mode amplitude, $y$ the spatial coordinate along the propagation direction, $y_{0}=3\,$µm the position of the wire termination, and $\tilde{\beta}=k_{0}\tilde{n}_\text{eff}$ the complex propagation constant with the effective mode index $n_\text{eff}=1.68+0.023i$ from the 2D mode analysis. We obtain the reflection amplitude $r=0.647$ and reflection phase $\phi_{r}=-1.89$.\\

\textbf{Collection efficiency}\\
We expect the far-field collection efficiencies for the scattered SPP and the direct QD emission to differ significantly, since the QD mostly radiates into the high-index GaAs substrate. Therefore, 3D finite element methods simulations are performed, approximating the QD as well as the out-coupled SPP as dipolar emitters, which are located 30\,nm below and 155\,nm above the semiconductor surface, respectively. The thickness of the spacer layer and the refractive indices are the same as in the 2D simulation in Fig.~1b. We use the RETOP package \cite{Yang2016} to obtain the amount of power which is radiated into the upper halfspace for the respective dipole positions. Normalized to the total emitted power, we find $\eta_{qd,ff}=0.026$ and $\eta_{spp,ff}=0.434$, leading to a far-field collection efficiency ratio $\frac{\eta_{spp,ff}}{\eta_{qd,ff}} \approx 17$.

\clearpage

\section{S6 - Influence of the QD dipole moment orientation}

\label{sec:dipoleorientation}

\begin{figure}[h]
\centering
\includegraphics[scale=1]{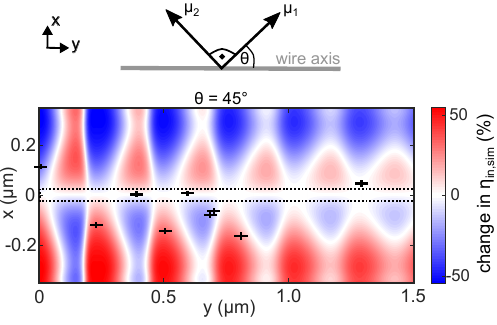}
\caption{\textbf{Influence of the dipole moment orientation on the coupling efficiency.}
Spatially resolved (relative) change in coupling efficiency for a hypothetical angle of $\theta=45^\circ$. The black crosses indicate the QD positions and the dashed lines the width of the nanowire.}
\label{fig:crossterm_theta}
\end{figure}

The general expression for the coupling efficiency of an emitter with dipole moment $\boldsymbol{\mu}$ in the modal field $\boldsymbol{E}$ of the waveguide is given by
\begin{equation}
\eta_{in} = |\boldsymbol{\mu} \cdot \boldsymbol{E} |^2,
\end{equation}
assuming normalized electric fields. Our epitaxial GaAs quantum dots feature two orthogonal excitonic states which both can be excited non-resonantly with our laser at $\lambda=635$\,nm. We assume that the corresponding dipole moment contributions $\boldsymbol{\mu}_{1}$ and $\boldsymbol{\mu}_{2}$ add up incoherently and the overall incoupling efficiency can be written as
\begin{equation}
\eta_{in} = |\boldsymbol{\mu}_{1} \cdot \boldsymbol{E} |^2 + |\boldsymbol{\mu}_{2} \cdot \boldsymbol{E} |^2.
\end{equation}
As the out-of-plane transition dipole moment $\mu_{z}=0$ is vanishing, we obtain
\begin{equation}
\eta_{in} = |\mu_{1,x} E_{x} + \mu_{1,y} E_{y}|^2 + |\mu_{2,x} E_{x} + \mu_{2,y} E_{y}|^2.
\end{equation}
Assuming $|\boldsymbol{\mu}_{1}|=|\boldsymbol{\mu}_{2}|$ and exploiting the orthogonality of the excitonic states leads to
\begin{equation}
\eta_{in} = |\mu \sin{\theta} E_{x} + \mu \cos{\theta} E_{y}|^2
+ |\mu \cos{\theta} E_{x} + \mu \sin{\theta} E_{y}|^2,
\end{equation}
with $\theta$ being the angle between dipole moment contribution $\boldsymbol{\mu}_{1}$ and wire axis. As the electric field is complex-valued, we use $|a+b|^2=|a|^2+|b|^2+2\,\Re(ab^{*})$ to finally obtain
\begin{equation}
\eta_{in} =
\mu^2 \left[ |E_{x}|^2 + |E_{y}|^2 + 4 \sin{\theta} \cos{\theta}\,\Re{(E_{x} E^{*}_{y})} \right ].
\label{eqn:crossterm}
\end{equation}
The crossterm describes the influence of the dipole moment orientation on the coupling efficiency and vanishes for angles $\theta=0^\circ$ and $\theta=90^\circ$, representing parallel and perpendicular orientation with respect to the nanowire, as assumed in the main text.  The spatial dependence of the modification in the coupling efficiency is shown in Fig.~\ref{fig:crossterm_theta} for $\theta=45^\circ$, where the crossterm in Eq.~\ref{eqn:crossterm} has its maximum. It can be seen that the dipole moment orientation is negligible close to the nanowire axis at $x=0$, where the electric field points in the propagation direction and the $E_{x}$-component is zero. For QDs far away from the nanowire axis, the coupling efficiency can be modified up to $\pm30\,\%$ at $\theta=45^\circ$. As the sign of the electric field is unkown, we take the absolute value of this hypothetical "worst case" modification as an additional uncertainty for the coupling efficiency of each QD in Fig.~3c in the main text.

\clearpage

\section{S7 - Propagation length}
\begin{figure}[h]
\centering
\includegraphics[scale=1]{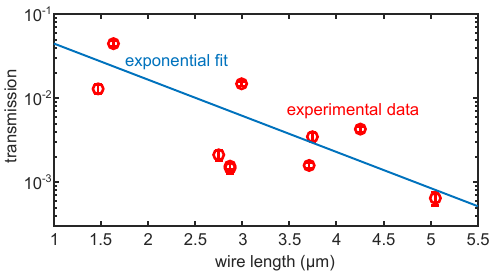}
\caption{\textbf{Laser transmission for several Ag nanowires to determine the surface plasmon propagation length.}
Red circles: measured SPP transmission. Blue line: exponential fit, suggesting a propagation length of $L_{p}\approx1.0$\,µm.}
\label{fig:Lprop}
\end{figure}

In order to determine the SPP propagation length, a modelocked Titanium-Sapphire laser is operated at a wavelength $\lambda=795$\,nm to match the QD emission wavelength. The laser is polarized parallel with respect to the nanowire axis and focused on one of its ends. The direct laser reflection as well as the out-coupled SPP emission at the other wire end is imaged onto the CCD-camera. In order to obtain the transmission (see Fig.~\ref{fig:Lprop}), the SPP emission is normalized to the laser reflection. The wire lengths are taken from SEM images, respectively. By fitting an exponential function on the transmission data, we find a \mbox{$1/e$-propagation length}~ $L_{p}\approx1.0$\,µm, consistent with the fit result in the main text. It must be stated that the spread of the data points is rather large, leading to a large uncertainty of the fit. One reason for that may be that the laser coupling is sensitive to the exact shape of the laser focus as well as lateral position regarding the nanowire end. In contrast, for the QD coupling experiment, the exact focusing conditions are less sensitive, as the laser only generates electron-hole pairs in the environment of the QD.

\bibliographystyle{achemso}

\end{document}